\documentclass[aps,prd,amsfonts,amssymb, amsmath, showpacs, showkeys, a4paper, nofootinbib, superscriptaddress, 11pt, raggedbottom]{revtex4-2}

\usepackage{braket}
\usepackage{bm}
\usepackage{graphicx}
\usepackage{slashed}
\usepackage{subfig}
\usepackage{mathrsfs}
\usepackage{color}
\usepackage{mathtools}
\usepackage[normalem]{ulem}
\usepackage{simplewick}
\usepackage{amsmath}

\newcommand{\kket}[1]{| #1 \rangle\!\rangle}
\newcommand{\bbra}[1]{\langle\!\langle #1 |}
\newcommand{\bbrakket}[1]{\langle\!\langle #1 \rangle\!\rangle}

\def\beq{\begin{equation}}
\def\eeq{\end{equation}}
\def\beqa{\begin{eqnarray}}
\def\eeqa{\end{eqnarray}}

\begin{document}

\title{\Large Density matrix formalism for interacting quantum fields}

\author{Christian K\"{a}ding}
\email{christian.kaeding@tuwien.ac.at}
\affiliation{Technische Universit\"at Wien, Atominstitut, Stadionallee 2, 1020 Vienna, Austria}

\author{Mario Pitschmann}
\email{mario.pitschmann@tuwien.ac.at}
\affiliation{Technische Universit\"at Wien, Atominstitut, Stadionallee 2, 1020 Vienna, Austria}

\begin{abstract}
We provide a description of interacting quantum fields in terms of density matrices for any occupation numbers in Fock space in a momentum basis. As a simple example, we focus on a real scalar field interacting with another real scalar field, and present a practicable formalism for directly computing the density matrix elements of the combined scalar-scalar system. For deriving the main formula, we use techniques from non-equilibrium quantum field theory like thermo field dynamics and the Schwinger-Keldysh formalism. Our results allow for studies of particle creation/annihilation processes at finite times and other non-equilibrium processes including those found in the theory of open quantum systems. 

\end{abstract}


\maketitle



\section{Introduction}
Density operators and matrices are powerful tools for describing the quantum state of a physical system since they not only allow for the description of pure states, but also of mixed ones. The latter is a feature not accessible with descriptions by wave functions \cite{Schlosshauer}. Consequently, besides their uses in non-relativistic quantum mechanics, density matrices also find ample applications in (open) quantum field theory \cite{Calabrese2004,Calabrese2005,Gavrilov2006,Calzetta2008,Casini2009,Balasubramanian2011,Boyanovsky_2015,Doyon2016,Sieberer2016,Marino2016,Baidya2017,Engelhardt2017,Burrage2018,Zhang2019,Dong2020,Nagy2020,Jana2021,Fogedby2022,Emonts2022,Kading2022_2,Boyanovsky2022,Kaplanek2022} and related areas, including cosmology \cite{Lombardo1,Lombardo2,Lombardo3,Boyanovsky1,Boyanovsky2,Boyanovsky3,Boyanovsky4,Burgess2015,Hollowood,Emond2018,Colas2021,Brahma2021, Brahma2022,Colas2022}, black holes \cite{Yu2008,Lombardo2012,Kaplanek2020,Burgess2021,Kaplanek2021} or heavy-ion physics \cite{Brambilla1,Brambilla2,Akamatsu2020,DeJong2020,Yao2021,Brambilla2021}. 
\\
The time evolution of a density operator is governed by the quantum Liouville equation, from which quantum master equations can be derived. Such master equations find heavy use in the theory of open quantum systems \cite{Breuer2002}. Especially in this context, it is common practice to trace or integrate out some degrees of freedom often referred to as environments. This process results in so-called reduced density matrices, which describe only the subsystems of the total system that were not traced out.   In Ref.\,\cite{Burrage2018} the authors introduced a Lehmann-Szymanzik-Zimmermann-like reduction (LSZ-like reduction) \cite{Lehmann1954} that provides a practicable and first principle-based formalism for deriving quantum master equations for reduced density matrices in a momentum basis in Fock space (also discussed in Refs.\,\cite{Burrage2019,Kading2019}). This was done with applications to quenched field theoretical open systems in mind. The formalism was then used in Ref.\,\cite{Kading2022_2} in order to provide a way of directly computing reduced density matrix elements for such systems in terms of Feynman-Vernon influence functionals \cite{Feynman} without having to solve potentially intricate quantum master equations.
\\
In the present article we will follow the path laid out by Ref.\,\cite{Kading2022_2}, but apply the formalism more generally to interacting quantum field theoretical systems without tracing out one of the degrees of freedom. Instead we will show how the formalism introduced in Ref.\,\cite{Burrage2018} can be used for computing total density matrix elements, which include information about all systems partaking in the described interactions, for any occupation in Fock space in a momentum basis. Having such a formalism can be a valuable tool, for example, when investigating decay or creation processes at finite times. Moreover, we see manifold direct applications at the high-precision frontier, i.e. in the area of low-energy but high-precision phenomenology as, e.g. in quantum optics experiments, which we will work on in the near future. The current article is a first step towards such applications.
\\
As a simple working example, we will focus our discussion on the interactions of real scalar fields. Extensions of the presented formalism to other types of fields are possible. For simplicity, for most of our discussion we will only consider the interaction between two scalar fields $\phi$ and $\varphi$. However, the formula that we will later present as a tool for directly computing their total density matrix elements will be extrapolated to a formula allowing to include any number of scalar field species.
\\
This article has the following structure: Sec.\,\ref{sec:Derivation} deals with the required mathematical concepts, which are then applied for deriving a formula for directly computing total density matrix elements for the combined system of $\phi$ and $\varphi$. A more general formula for any number of scalar field species will also be extrapolated. Subsequently, in Sec.\,\ref{sec:Example}, a few examples are investigated using the derived formula before the article is concluded in Sec.\,\ref{sec:Conclusion}.


\section{Derivation}
\label{sec:Derivation}

In this section we will derive a formula for the direct computation of total density matrices in a momentum basis for any number of particles and species of interacting quantum fields. Since they constitute the simplest example, we will restrict our discussion to real scalar fields. However, similar formalisms can be developed for other types of fields as well. At first, we will only consider two interacting scalar fields $\phi$ and $\varphi$, but generalising the resulting equation to more scalar fields is straightforward.  
\\
Initially, we will introduce all required mathematics and concepts in Secs.\,\ref{ssec:Density}-\ref{ssec:TFD}. For this, we will closely follow Ref.\,\cite{Kading2022_2}, which in turn is strongly based on the discussions in Ref.\,\cite{Burrage2018}. Finally, we will present the derivation of a formula for the direct computation of total density matrices in Sec.\,\ref{ssec:EQN}.


\subsection{Density matrices in Fock space}
\label{ssec:Density}

Since we are interested in describing the interaction of quantum fields in terms of density matrices, we first need to introduce them in Fock space. For this, we expand the density operator in a momentum basis similarly to Ref.\,\cite{Kading2022_2}, but while taking into account multiple numbers of fields: 
\begin{eqnarray}\label{eq:DOFock}
\hat{\rho}(t) &=& \sum\limits_{I,J=0}^\infty \frac{1}{I!J!}\int\left(   \prod\limits_{A = 1}^I d\Pi_{K^A}\right)\left(\prod\limits_{B = 1}^J   d\Pi_{L^B} \right)\rho_{I;J}(K^I|L^J|t) \ket{K^I}\bra{L^J}\,\,\,,\,\,\,\,\,\,\,\,\,\,\,
\end{eqnarray}
where we defined multi-indices 
\begin{eqnarray}
I &:=& (i_\alpha, i_\beta,...)\,\,\,,\,\,\,\,\,\,
J \,:=\, (j_\alpha, j_\beta,...)
\end{eqnarray}
with $\alpha$, $\beta$,... representing different field or particle species, such that
\begin{eqnarray}
I! &:=& i_\alpha ! i_\beta !...\,\,\,.
\end{eqnarray}
Notice that the cases $i_\alpha=0$ or $j_\alpha=0$ correspond to the static vacuum state $\ket{0}$ or $\bra{0}$ for the respective species $\alpha$. If $I =0$ or $J=0$, then this describes the case of no particles being present, i.e. the total vacuum.
Furthermore, we introduced the short-hand notation
\begin{eqnarray}
K^I &:=& \mathbf{k}_\alpha^{(1)},...,\mathbf{k}_\alpha^{(i_\alpha)};\mathbf{k}_\beta^{(1)},...,\mathbf{k}_\beta^{(i_\beta)};...
\,\,\,,\,\,\,\,\,\,
L^J \,:=\, \mathbf{l}_\alpha^{(1)},...,\mathbf{l}_\alpha^{(j_\alpha)};\mathbf{l}_\beta^{(1)},...,\mathbf{l}_\beta^{(j_\beta)};...
\end{eqnarray}
for the $3$-momenta, and 
\begin{eqnarray}
\prod\limits_{A = 1}^I d\Pi_{K^A} &:=& \left(\prod\limits_{a = 1}^{i_\alpha} d\Pi_{\mathbf{k}_\alpha^{(a)}}\right)\left(\prod\limits_{a = 1}^{i_\beta} d\Pi_{\mathbf{k}^{(a)}_\beta}\right)... \,\,\,,
\end{eqnarray}
where 
\begin{eqnarray}
\int d\Pi_{\mathbf{k}_\alpha} &:=& \int_{\mathbf{k}_\alpha} \frac{1}{2E_{\mathbf{k}_\alpha}^\alpha}
\end{eqnarray}
with\footnote{Later we will also make use of $\int_{k} := \int \frac{d^4k}{(2\pi)^4}$ for the $4$-momenta.}
\begin{eqnarray}
\int_{\mathbf{k}} &:=& \int \frac{d^3k}{(2\pi)^3}\,\,\,.
\end{eqnarray}
The density matrix elements that appear in Eq.\,(\ref{eq:DOFock}) are given by
\begin{eqnarray}\label{eq:DensMatrixdef}
\rho_{I;J}(K^I|L^J|t) &=& \bra{K^I}\hat{\rho}(t)\ket{ L^J}\,\,\,,
\end{eqnarray}
and, as usual, have to fulfill:
\begin{eqnarray}\label{eq:DensProperty}
\rho_{I;J}(K^I|L^J|t) &=& \rho^\ast_{J;I}(L^J|K^I|t)\,\,\,.
\end{eqnarray}
In addition, as was pointed out in Refs.\,\cite{Millington:2012pf,Millington:2013isa}, density matrix elements are picture-independent.
We physically interpret them in the following way: $\rho_{0;0}$ describes a $0$-particle or vacuum state, $\rho_{(i_\alpha,...);0}$ or $\rho_{0;(j_\alpha,...)}$ stand for correlations between $i_\alpha$- or $j_\alpha$-$\alpha$-particles with the vacuum, while $\rho_{(i_\alpha,...);(j_\alpha,...)}$ represent correlations between $i_\alpha$- and $j_\alpha$-$\alpha$-particle states. 
\\
For later reference, it should also be noted that, in the Schr\"odinger picture (index $S$), the time evolution of a density operator is determined by the quantum Liouville equation\footnote{We use $\hbar \equiv 1$ throughout the entire article.} \cite{Breuer2002}
\begin{eqnarray}\label{Eqn:Liouville}
\frac{\partial}{\partial t} \hat{\rho}_S(t) &=& -\mathrm{i}[\hat{H}_S(t),\hat{\rho}_S(t)]\,\,\,,
\end{eqnarray}
which is generally solved by
\begin{eqnarray}\label{eqn:SPS2}
\hat{\rho}_S(t) &=& (\mathrm{T} e^{-\mathrm{i}\int^t_0 d\tau \hat{H}_S(\tau)})\hat{\rho}(0)(\tilde{\mathrm{T}} e^{\mathrm{i}\int^t_0 d\tau \hat{H}_S(\tau)})\,\,\, 
\end{eqnarray}
if the Hamiltonian is time-dependent due to an external source.


\subsection{The Schwinger-Keldysh formalism}
\label{ssec:FVIF}

\begin{figure}[htbp]
\begin{center}
\includegraphics[scale=0.5]{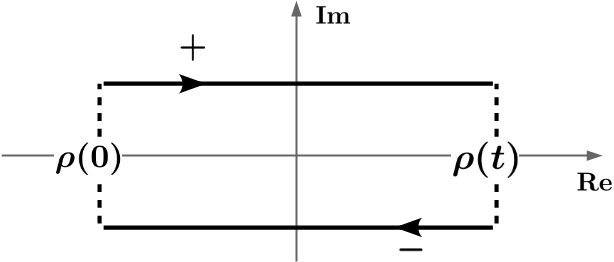}
\caption{\textit{(Taken from Ref.\,\cite{Kading2022_2})} Schematic depiction of the closed time path for a density matrix $\rho$ evolving from an initial time $0$ to a final time $t$ and backwards}
\label{fig:CTP}
\end{center}
\end{figure}

When computing expectation values for operators at finite times, for example the density matrix elements in Eq.\,(\ref{eq:DensMatrixdef}), the usual in-out formalism known from scattering amplitudes is not sufficient anymore. Instead, the Schwinger-Keldysh closed-time-path formalism \cite{Schwinger,Keldysh}, also known as in-in formalism, is used. A good introduction to this topic can, for example, be found in Ref.\,\cite{Calzetta2008}. The Schwinger-Keldysh formalism is essentially relying on doubling the degrees of freedom, where the two copies are distinguished by labels $+$/$-$, and letting those evolve on the positive/negative branch of the closed time path depicted in Fig.\,\ref{fig:CTP} between an initial $t_\text{initial} =0$ and a final time $t_\text{final} =t$ \cite{Kading2022_2}. While for scattering amplitudes in- and out-states are used, which are taken to be asymptotic, the path-integral representation of the trace of an operator at a finite time requires us to consider two copies of field-states and therefore naturally leads to the closed time path. This formalism is often applied in the context of field theoretical descriptions of open quantum systems, e.g.\,\,in Ref.\,\cite{Burrage2018}, in order to define the so-called Feynman-Vernon influence functional \cite{Feynman}. 
If we consider the example of two real scalar fields $\phi$ and $\varphi$ interacting with each other, then, within the the Schwinger-Keldysh formalism, we can define the density functional\footnote{The extension to more field species is straightforward.} 
\begin{eqnarray}
\rho[\phi^\pm_t;\varphi^\pm_t|t] &:=& 
\langle\phi^+_t,\varphi^+_t|\hat{\rho}(t)|\phi^-_t, \varphi^-_t\rangle\,\,\,,
\end{eqnarray}
where $\pm$ indicates a dependence on both $+$- and $-$-type operators, and subscript $t$ labels the time slice on which the field eigenstates have to be taken \cite{Kading2022_2}. The time evolution of such a density functional 
\begin{eqnarray}\label{eq:DensFunctional}
\rho[\phi^\pm_t;\varphi^\pm_t|t] &=& \int d\phi_0^\pm d\varphi_0^\pm \mathcal{P}[\phi^\pm_t,\phi^\pm_0;\varphi^\pm_t,\varphi^\pm_0|t,0] \rho[\phi^\pm_0;\varphi^\pm_0|0]
\end{eqnarray}
is given in terms of the functional propagator\footnote{Compare it to the influence functional propagator in the literature, e.g. in Ref.\,\,\cite{Calzetta2008}.},
\begin{eqnarray}\label{Eqn:IFProp}
\mathcal{P}[\phi^\pm_t,\phi^\pm_0;\varphi^\pm_t,\varphi^\pm_0|t,0] &=& \int^{\phi^\pm_t}_{\phi^\pm_0}\mathcal{D}\phi^\pm \int^{\varphi^\pm_t}_{\varphi^\pm_0} \mathcal{D}\varphi^\pm e^{\mathrm{i}\widehat{S}[\phi;\varphi|t]}\,\,\,
\end{eqnarray}
with the action 
\begin{eqnarray}\label{Eqn:EffS}
\widehat{S}[\phi;\varphi|t] &=& \widehat{S}_\phi[\phi|t] + \widehat{S}_\varphi[\varphi|t] + \widehat{S}_{\phi,\text{int}}[\phi|t]+ \widehat{S}_{\varphi,\text{int}}[\varphi|t]+ \widehat{S}_\text{int}[\phi;\varphi|t]\,\,\,.
\end{eqnarray}
Here, ~$\widehat{}$~ indicates functionals that depend on both field variables $\phi^+$ and $\phi^-$, such that
\begin{eqnarray}\label{eq:PMActions}
\widehat{S}[\phi;\varphi|t] &:=& S[\phi^+;\varphi^+|t] - S[\phi^-;\varphi^-|t]\,\,\,.
\end{eqnarray}
The action $\widehat{S}_\alpha[\alpha|t]$ is for a free $\alpha$-field, $\widehat{S}_{\alpha,\text{int}}[\alpha|t]$ is the corresponding self-interaction action, and $\widehat{S}_\text{int}[\phi;\varphi|t]$ is the action describing the interaction between the fields $\phi$ and $\varphi$.
Since we are only interested in finite times, we restrict all actions to $\Omega_t := [0,t]\times\mathbb{R}^3$, such that, schematically, we use 
\begin{eqnarray}
S[t]  &=& \int_{x\in\Omega_t} \mathcal{L}[x]
\end{eqnarray}
with 
\begin{eqnarray}
\int_{x\in\Omega_t} &:=& \int_{\Omega_t} d^4x\,\,\,.
\end{eqnarray}
Furthermore, notice the notational distinction between functional integrals over all of $\mathbb{R}^3$ on a particular time slice, denoted by $d$ as in Eq.\,(\ref{eq:DensFunctional}), and over the whole $\Omega_t$, denoted by $\mathcal{D}$ as in Eq.\,(\ref{Eqn:IFProp}).


\subsection{Thermo field dynamics}
\label{ssec:TFD}

The final concept we need to discuss is thermo field dynamics (TFD) \cite{Takahasi:1974zn,Arimitsu:1985ez,Arimitsu:1985xm} (see also Ref.\,\cite{Khanna}). For this, we will directly quote from most of the corresponding elaborations in Ref.\,\cite{Kading2022_2}. One way of understanding TFD is as an algebraic formulation of the Schwinger-Keldysh formalism, in the sense that it works with operators on a positive $(+)$ or negative $(-)$ complex time axis acting on a doubled Hilbert space
\begin{eqnarray}
\widehat{\mathcal{H}} &:=& \mathcal{H}^+ \otimes \mathcal{H}^-\,\,\,,
\end{eqnarray} 
where $\mathcal{H}^\pm$ are the Hilbert spaces corresponding to the $\pm$-branches of the closed time path.
Operators living on the closed time path can be expressed within TFD as
\begin{eqnarray}
\hat{\mathcal{O}}^+ &=& \hat{\mathcal{O}} \otimes \hat{\mathbb{I}}\,\,\,,
\nonumber
\\
\hat{\mathcal{O}}^- &=& \hat{\mathbb{I}} \otimes \hat{\mathcal{O}}^\mathcal{T}\,\,\,,
\end{eqnarray}
where $\hat{\mathbb{I}}$ is the unit operator and $\mathcal{T}$ means time reversal.
States in a momentum basis in TFD can be reached from the doubled vacuum state
\begin{eqnarray}
\kket{0} &:=& \ket{0} \otimes \ket{0}
\end{eqnarray}
via creation operators
\begin{eqnarray}\label{eq:Creators}
\hat{a}^{+\dagger}_\mathbf{k} \kket{0} \,=\, \ket{\mathbf{k}} \otimes \ket{0} \,=:\, \kket{\mathbf{k}_+}
\,\,\,,\,\,\,\,\,\,\,\,\,
\hat{a}^{-\dagger}_\mathbf{k} \kket{0} \,=\, \ket{0} \otimes \ket{\mathbf{k}} \,=:\, \kket{\mathbf{k}_-}\,\,\,.
\end{eqnarray}
Consequently, the corresponding annihilators act like
\begin{eqnarray}\label{eq:Annihilators}
\hat{a}^{\pm}_\mathbf{k} \kket{\mathbf{p}_+,\mathbf{p}_-} \,=\, (2\pi)^3 2 E^\phi_\mathbf{k}\delta^{(3)}(\mathbf{p}-\mathbf{k})\kket{\mathbf{p}_\mp}
\,\,\,.
\end{eqnarray}
In addition, we can construct a special state corresponding to the unit operator, see Ref.\,\cite{Arimitsu:1985ez},
\begin{eqnarray}\label{eq:SpecState}
\kket{1} &:=& \kket{0} + \int d\Pi_{\mathbf{k}} \kket{\mathbf{k}_+,\mathbf{k}_-} + \frac{1}{2!}\int d\Pi_{\mathbf{k}} d\Pi_{\mathbf{k'}}\kket{\mathbf{k}_+,\mathbf{k'}_+,\mathbf{k}_-,\mathbf{k'}_-}+...\,\,\,,
\end{eqnarray}
which allows us to express the expectation value of an operator as
\begin{eqnarray}\label{eq:TFDtrace}
\braket{\hat{\mathcal{O}}(t)} &=& \text{Tr}\hat{\mathcal{O}}(t)\hat{\rho}(t) \,=\, \bbrakket{1| \hat{\mathcal{O}}^+(t)\hat{\rho}^+(t) |1}\,\,\,.
\end{eqnarray}
Having all this, we can now use the fact that Eq.\,(\ref{Eqn:Liouville}) can be re-written in a Schr\"odinger-like form:
\begin{eqnarray}\label{Eqn:S-likeLiou}
\frac{\partial}{\partial t} \hat{\rho}_S^+(t)\kket{1}_S &=& -\mathrm{i}\widehat{H}_S(t)\hat{\rho}_S^+(t)\kket{1}_S\,\,\,,
\end{eqnarray}
where $\widehat{H}_S(t) := \hat{H}_S(t) \otimes \mathbb{I} - \mathbb{I} \otimes \hat{H}_S(t)$. This can generally be solved by
\begin{eqnarray}\label{eq:SolutionTFD}
\hat{\rho}_S^+(t)\kket{1}_S &=& \text{T} \exp\left\{{-\mathrm{i}\int\limits_{0}^t\widehat{H}_S(\tau)d\tau}\right\}\hat{\rho}^+(0)\kket{1}_S\,\,\,.
\end{eqnarray}
Note that at time $0$ the different pictures coincide and we therefore dropped the label $S$.
\\
We recall that, generally, Hamiltonian and action are related via $H(t) = -\frac{\partial}{\partial t} S(t)$. Furthermore, if we consider that the action in Eq.\,(\ref{Eqn:EffS}) describes the full evolution of the density matrix elements in the field-basis, the interaction picture $\widehat{H}_{I}$ must be its corresponding Hamiltonian, such that $\hat{H}_{I}(t) := \hat{H}_{0,I}(t) + \hat{H}_{\phi,\text{int},I}(t) + \hat{H}_{\varphi,\text{int},I}(t) + \hat{H}_{\text{int},I}(t)$ with $\hat{H}_{0,I}(t) := \hat{H}_{\phi,0,I}(t) + \hat{H}_{\varphi,0,I}(t)$ representing the action $S[\phi;\varphi|t]$ as appearing in Eq.\,(\ref{eq:PMActions}). This will later be of greater importance when we write down a path integral expression for the density matrix elements.
For now, however, we just remind ourselves that the free Hamiltonian $\hat{H}_0$ is the same in Schr\"odinger and interaction picture, and consequently drop the subscript.
\\
Translating Eq.\,(\ref{Eqn:S-likeLiou}) into the interaction picture, and using the fact that the state $\kket{1}$ is actually time- and picture-independent \cite{Kading2022_2}, therefore leaves us with 
\begin{eqnarray}
 \partial_t\hat{\rho}_{I}^+(t)\kket{1} &=& -\mathrm{i}[\widehat{H}_{\phi,\text{int},I}(t) + \widehat{H}_{\varphi,\text{int},I}(t) + \widehat{H}_{\text{int},I}(t)]\hat{\rho}_{I}^+(t)\kket{1}\,\,\,,
\end{eqnarray}
which can be solved by
\begin{eqnarray}\label{eq:AltSol}
\hat{\rho}_{I}^+(t)\kket{1}
&=& \text{T} \exp\left\{{-\mathrm{i}\int\limits_{0}^t[\widehat{H}_{\phi,\text{int},I}(\tau)+ \widehat{H}_{\varphi,\text{int},I}(\tau)+\widehat{H}_{\text{int},I}(\tau)]d\tau}\right\}\hat{\rho}^+(0)\kket{1}
\,\,\,.
\end{eqnarray}


\subsection{Density matrix elements}
\label{ssec:EQN}

We can now use the concepts discussed in the previous sections in order to derive an equation for the direct computation of elements of the total density matrix for two interacting real scalar fields $\phi$ and $\varphi$ in a momentum basis for any occupation number in Fock space. Doing so, we will only work in the interaction picture and therefore not use any corresponding labels for operators and states anymore.
\\
Initially, we will derive a formula that allows for the computation of the density matrix element $\rho_{1,1;1,1}(\mathbf{p};\mathbf{k}|\mathbf{p}';\mathbf{k}' |t)$, which represents the state of a single $\phi$-particle (momenta $\mathbf{p}$ and $\mathbf{p}'$) and a single $\varphi$-particle (momenta $\mathbf{k}$ and $\mathbf{k}'$). Having found this expression, it will later be possible to extrapolate a more general formula for any number of particle species and respective occupation numbers in momentum space.
\\
As in Refs.\,\cite{Burrage2018,Kading2022_2}, our starting point is 
\begin{eqnarray}\label{eq:335}
\langle\mathbf{p};\mathbf{k};t|\hat{\rho}(t)|\mathbf{p}';\mathbf{k}';t\rangle &=& \rho_{1,1;1,1}(\mathbf{p};\mathbf{k}|\mathbf{p}';\mathbf{k}' |t)\,\,\,.
\end{eqnarray}
Using Eq.\,(\ref{eq:TFDtrace}), Eq.\,(\ref{eq:335}) can be rewritten as
\begin{eqnarray}
\text{Tr}|\mathbf{p}';\mathbf{k}';t\rangle\langle\mathbf{p};\mathbf{k};t|\hat{\rho}(t) &=& \langle\langle 1|(|\mathbf{p}';\mathbf{k}';t\rangle \langle\mathbf{p};\mathbf{k};t|\otimes \hat{\mathbb{I}})\hat{\rho}^+(t) |1\rangle\rangle\,\,\,
\end{eqnarray}  
in TFD language. Next, we substitute Eq.\,(\ref{eq:AltSol}) and find
\begin{eqnarray}\label{eq:zwiSchr1}
\rho_{1,1;1,1}(\mathbf{p};\mathbf{k} |\mathbf{p}';\mathbf{k}'|t) &=&  \langle\langle \mathbf{p}_+,\mathbf{p}'_-;\mathbf{k}_+,\mathbf{k}'_-;t| 
\nonumber
\\
&&\,\,\,\,\,\,
\times
\text{T}\exp\left\{{-\mathrm{i}\int\limits_{0}^t[\widehat{H}_{\phi,\text{int}}(\tau)+ \widehat{H}_{\varphi,\text{int}}(\tau)+\widehat{H}_{\text{int}}(\tau)]d\tau}\right\}\hat{\rho}^+(0) |1\rangle\rangle\,\,\,.\,\,\,\,\,\,\,\,\,\,\,\,
\end{eqnarray}
In order to proceed, we need to expand the density operator on the right-hand side of Eq.\,(\ref{eq:zwiSchr1}) as in Eq.\,(\ref{eq:DOFock}). However, in this way we would encounter the problem of potentially having to deal with an infinite number of initial density matrix elements. Luckily, we can make the in many experimental situations reasonable assumption  that we know the initial setup sufficiently well, such that we can ignore all vanishing or suppressed occupations. In our particular example, we will assume that the total initial system already consists of only one $\phi$- and one $\varphi$-particle. Consequently, we only have to consider the initial density matrix element for the case $i_\phi=i_\varphi=j_\phi=j_\varphi=1$ and assume all others to be nil. We do this assumption for the sake of readability, but a more general result will later be presented, which could be obtained by modifying the present derivation accordingly. Under the considered assumptions we are led to
\begin{eqnarray}
\rho_{1,1;1,1}(\mathbf{p};\mathbf{k} |\mathbf{p}';\mathbf{k}'|t) &=&  \langle\langle \mathbf{p}_+,\mathbf{p}'_-;\mathbf{k}_+,\mathbf{k}'_-;t| \text{T} \exp\left\{{-\mathrm{i}\int\limits_{0}^t[\widehat{H}_{\phi,\text{int}}(\tau)+ \widehat{H}_{\varphi,\text{int}}(\tau)+\widehat{H}_{\text{int}}(\tau)]d\tau}\right\}
\nonumber
\\
&&\,\,\,\,\,\,
\times
\int d\Pi_{\mathbf{q}}d\Pi_{\mathbf{q}'}d\Pi_{\mathbf{l}}d\Pi_{\mathbf{l}'}\rho_{1,1;1,1}(\mathbf{q};\mathbf{l}|\mathbf{q}';\mathbf{l}'|0) |\mathbf{q}_+,\mathbf{q}'_-;\mathbf{l}_+,\mathbf{l}'_-\rangle\rangle\,\,\,,\,\,\,\,\,\,\,\,\,\,\,\,\,\,\,
\end{eqnarray}
which, using Eqs.\,(\ref{eq:Creators}) and (\ref{eq:Annihilators}) becomes
\begin{eqnarray}
\rho_{1,1;1,1}(\mathbf{p};\mathbf{k} |\mathbf{p}';\mathbf{k}'|t) &=&  \int d\Pi_{\mathbf{q}}d\Pi_{\mathbf{q}'}d\Pi_{\mathbf{l}}d\Pi_{\mathbf{l}'}\rho_{1,1;1,1}(\mathbf{q};\mathbf{l}|\mathbf{q}';\mathbf{l}'|0) \langle\langle 0|\text{T}\hat{a}^+_{\mathbf{p}}(t) \hat{a}^-_{\mathbf{p} '}(t)\hat{b}^+_{\mathbf{k}}(t) \hat{b}^-_{\mathbf{k} '}(t)
\nonumber
\\
&&\,\,\,\,\,\,
\times
\exp\left\{{-\mathrm{i}\int\limits_{0}^t[\widehat{H}_{\phi,\text{int}}(\tau)+ \widehat{H}_{\varphi,\text{int}}(\tau)+\widehat{H}_{\text{int}}(\tau)]d\tau}\right\}
\nonumber
\\
&&\,\,\,\,\,\,
\times
\hat{a}^{+\dagger}_{\mathbf{q}}(0) \hat{a}^{-\dagger}_{\mathbf{q} '}(0)\hat{b}^{+\dagger}_{\mathbf{l}}(0) \hat{b}^{-\dagger}_{\mathbf{l} '}(0)|0\rangle\rangle\,\,\,.\,\,\,\,\,\,\,
\end{eqnarray}
Here we denoted creators/annihilators for $\phi$ with $\hat{a}^\dagger$/$\hat{a}$ and those for $\varphi$ with $\hat{b}^\dagger$/$\hat{b}$.
Next, we replace them by (cf.\,\,Ref.\,\cite{Burrage2018})
\begin{eqnarray}
\hat{a}^+_{\mathbf{p}}(t) &=& +\mathrm{i}\int_{\mathbf{x}} e^{-\mathrm{i}\mathbf{p}\cdot\mathbf{x}}\partial_{t,E^\phi_{\mathbf{p}}}\hat{\phi}^+(t,\mathbf{x})\,\,\,,\,\,\,\,\,\,\,\,\,
\hat{a}^{+\dagger}_{\mathbf{p}}(t) \,=\, -\mathrm{i}\int_{\mathbf{x}} e^{+\mathrm{i}\mathbf{p}\cdot\mathbf{x}} \partial_{t,E^\phi_{\mathbf{p}}}^*\hat{\phi}^+(t,\mathbf{x})\,\,\,,
\nonumber
\\
\hat{a}^-_{\mathbf{p}}(t) &=& -\mathrm{i}\int_{\mathbf{x}} e^{+\mathrm{i}\mathbf{p}\cdot\mathbf{x}}\partial_{t,E^\phi_{\mathbf{p}}}^*\hat{\phi}^-(t,\mathbf{x})\,\,\,,\,\,\,\,\,\,\,\,\,
\hat{a}^{-\dag}_{\mathbf{p}}(t) \,=\, +\mathrm{i}\int_{\mathbf{x}} e^{-\mathrm{i}\mathbf{p}\cdot\mathbf{x}}\partial_{t,E^\phi_{\mathbf{p}}}\hat{\phi}^{-}(t,\mathbf{x})\,\,\,,
\end{eqnarray}
where $\partial_{t,E^\phi_{\mathbf{p}}} := \overset{\rightarrow}{\partial}_t - \mathrm{i}E^\phi_{\mathbf{p}}$, and there exist corresponding expressions for $\hat{b}^\dagger$/$\hat{b}$.
This leaves us with
\begin{eqnarray}\label{eq:FinOpEx}
\rho_{1,1;1,1}(\mathbf{p};\mathbf{k} |\mathbf{p}';\mathbf{k}'|t) &=&  \lim_{\substack{x^{0(\prime)}_{\phi,\varphi}\,\to\, t^{+}\\y^{0(\prime)}_{\phi,\varphi}\,\to\, 0^-}}
\int d\Pi_{\mathbf{q}}d\Pi_{\mathbf{q}'}d\Pi_{\mathbf{l}}d\Pi_{\mathbf{l}'}\rho_{1,1;1,1}(\mathbf{q};\mathbf{l}|\mathbf{q}';\mathbf{l}'|0)
\nonumber
\\
&&\,\,\,\,\,\,
\times
\left(\prod_{a=\phi,\varphi}
\int_{\mathbf{x}_a\mathbf{x}'_a\mathbf{y}_a\mathbf{y}'_a} \right)
e^{-\mathrm{i}(\mathbf{p}\cdot\mathbf{x}_\phi-\mathbf{p}'\cdot\mathbf{x}'_\phi)+\mathrm{i}(\mathbf{q}\cdot\mathbf{y}_\phi-\mathbf{q}'\cdot\mathbf{y}'_\phi)}e^{-\mathrm{i}(\mathbf{k}\cdot\mathbf{x}_\varphi-\mathbf{k}'\cdot\mathbf{x}'_\varphi)+\mathrm{i}(\mathbf{l}\cdot\mathbf{y}_\varphi-\mathbf{l}'\cdot\mathbf{y}'_\varphi)}
\nonumber
\\
&&\,\,\,\,\,\,
\times
\partial_{x_\phi^0,E^\phi_{\mathbf{p}}}\partial_{x^{0\prime}_\phi,E^\phi_{\mathbf{p}'}}^*\partial_{y^0_\phi,E^\phi_{\mathbf{q}}}^*\partial_{y^{0\prime}_\phi,E^\phi_{\mathbf{q}'}}
\partial_{x^0_\varphi,E^\varphi_{\mathbf{k}}}\partial_{x^{0\prime}_\varphi,E^\varphi_{\mathbf{k}'}}^*\partial_{y^0_\varphi,E^\varphi_{\mathbf{l}}}^*\partial_{y^{0\prime}_\varphi,E^\varphi_{\mathbf{l}'}}
\nonumber
\\
&&\,\,\,\,\,\,
\times
\langle\langle 0|{\rm T}[\hat{\phi}^+(x_\phi)\hat{\phi}^-(x'_\phi)\hat{\varphi}^+(x_\varphi)\hat{\varphi}^-(x'_\varphi)
\nonumber
\\
&&\,\,\,\,\,\,
\times
\exp\left\{{-\mathrm{i}\int\limits_{0}^t[\widehat{H}_{\phi,\text{int}}(\tau)+ \widehat{H}_{\varphi,\text{int}}(\tau)+\widehat{H}_{\text{int}}(\tau)]d\tau}\right\}
\nonumber
\\
&&\,\,\,\,\,\,
\times
\hat{\phi}^+(y_\phi)\hat{\phi}^-(y'_\phi)\hat{\varphi}^+(y_\varphi)\hat{\varphi}^-(y'_\varphi)]|0\rangle\rangle\,\,\,.\,\,\,\,\,
\end{eqnarray}
As in Ref.\,\cite{Kading2022_2}, we introduced limits in which $x_{\phi,\varphi}^{0(\prime)}$ approach $t$ from above and $y_{\phi,\varphi}^{0(\prime)}$ approach $0$ from below in order to recover the correct time ordering.
\\
We translate the expression in Eq.\,(\ref{eq:FinOpEx}) into the path integral formalism, which yields
\begin{eqnarray}\label{eqn:11DensGen}
\rho_{1,1;1,1}(\mathbf{p};\mathbf{k} |\mathbf{p}';\mathbf{k}'|t) &=&  \lim_{\substack{x^{0(\prime)}_{\phi,\varphi}\,\to\, t^{+}\\y^{0(\prime)}_{\phi,\varphi}\,\to\, 0^-}}
\int d\Pi_{\mathbf{q}}d\Pi_{\mathbf{q}'}d\Pi_{\mathbf{l}}d\Pi_{\mathbf{l}'}\rho_{1,1;1,1}(\mathbf{q};\mathbf{l}|\mathbf{q}';\mathbf{l}'|0)
\nonumber
\\
&&\,\,\,\,\,\,
\times
\left(\prod_{a=\phi,\varphi}
\int_{\mathbf{x}_a\mathbf{x}'_a\mathbf{y}_a\mathbf{y}'_a} \right)
e^{-\mathrm{i}(\mathbf{p}\cdot\mathbf{x}_\phi-\mathbf{p}'\cdot\mathbf{x}'_\phi)+\mathrm{i}(\mathbf{q}\cdot\mathbf{y}_\phi-\mathbf{q}'\cdot\mathbf{y}'_\phi)}e^{-\mathrm{i}(\mathbf{k}\cdot\mathbf{x}_\varphi-\mathbf{k}'\cdot\mathbf{x}'_\varphi)+\mathrm{i}(\mathbf{l}\cdot\mathbf{y}_\varphi-\mathbf{l}'\cdot\mathbf{y}'_\varphi)}
\nonumber
\\
&&\,\,\,\,\,\,
\times
\partial_{x_\phi^0,E^\phi_{\mathbf{p}}}\partial_{x^{0\prime}_\phi,E^\phi_{\mathbf{p}'}}^*\partial_{y^0_\phi,E^\phi_{\mathbf{q}}}^*\partial_{y^{0\prime}_\phi,E^\phi_{\mathbf{q}'}}
\partial_{x^0_\varphi,E^\varphi_{\mathbf{k}}}\partial_{x^{0\prime}_\varphi,E^\varphi_{\mathbf{k}'}}^*\partial_{y^0_\varphi,E^\varphi_{\mathbf{l}}}^*\partial_{y^{0\prime}_\varphi,E^\varphi_{\mathbf{l}'}}
\nonumber
\\
&&\,\,\,\,\,\,
\times
\int\mathcal{D}\phi^{\pm}\mathcal{D}\varphi^{\pm} e^{\mathrm{i}\widehat{S}_0[\phi;\varphi|t]}\phi^+(x_\phi)\phi^-(x'_\phi)\varphi^+(x_\varphi)\varphi^-(x'_\varphi)
\nonumber
\\
&&\,\,\,\,\,\,
\times
\exp\left\{\mathrm{i}[\widehat{S}_{\phi,\text{int}}[\phi;t]+ \widehat{S}_{\varphi,\text{int}}[\varphi;t] +\widehat{S}_\text{int}[\phi;\varphi;t]]\right\}\phi^+(y_\phi)\phi^-(y'_\phi)\varphi^+(y_\varphi)\varphi^-(y'_\varphi)
\,\,\,,
\nonumber
\\
\end{eqnarray}
where we defined
\begin{eqnarray}
\widehat{S}_0[\phi;\varphi|t] &:=& \widehat{S}_{\phi}[\phi]+\widehat{S}_{\varphi}[\varphi]
\,\,\,,
\end{eqnarray}
and used 
\begin{eqnarray}
\widehat{H}_{\phi,\text{int}}(\tau)+ \widehat{H}_{\varphi,\text{int}}(\tau)+\widehat{H}_{\text{int}}(\tau) &=& \widehat{H}(\tau) - \widehat{H}_0(\tau)
\nonumber
\\
&=&
\frac{\partial}{\partial \tau} \widehat{S}_0(\tau) - \frac{\partial}{\partial \tau} \widehat{S}(\tau)
\nonumber
\\
&=&
\frac{\partial}{\partial \tau}\widehat{S}_0(\tau) + \sum_{a=\pm} a(\dot{\phi}^a \pi^a_\phi + \dot{\varphi}^a \pi^a_\varphi) - \left[\frac{\partial}{\partial \tau} \widehat{S}(\tau) + \sum_{a=\pm} a(\dot{\phi}^a \pi^a_\phi + \dot{\varphi}^a \pi^a_\varphi)\right]
\nonumber
\\
&=&
\frac{d}{d \tau} \widehat{S}_0(\tau) - \frac{d}{d \tau} \widehat{S}(\tau)
\nonumber
\\
&=&
-\frac{d}{d \tau}[\widehat{S}_{\phi,\text{int}}(\tau)+ \widehat{S}_{\varphi,\text{int}}(\tau) +\widehat{S}_\text{int}(\tau)]
\,\,\,.
\end{eqnarray}
The Eq.\,(\ref{eqn:11DensGen}) we just found allows us to directly compute the density matrix element that describes the interaction between two single particles $\phi$ and $\varphi$. Taking Eq.\,(\ref{eqn:11DensGen}), we extrapolate this formula for a general density matrix element with any number of particle species and occupations in Fock space. It can easily be seen that this result can be derived by extending the derivation for Eq.\,(\ref{eqn:11DensGen}) accordingly. We find:
\begin{eqnarray}\label{eq:GenDensForm}
\rho_{G;H}(K^G|L^H|t)
&=&
\sum\limits_{I,J=0}^\infty \frac{\mathrm{i}^{G+J}(-\mathrm{i})^{H+I}}{I!J!}
\lim_{\substack{X^{0,G},X^{0\prime,H}\,\to\, t^{+}\\Y^{0,I},Y^{0\prime,J}\,\to\, 0^-}}
\int \left(   \prod\limits_{A = 1}^I d\Pi_{R^A}\right)\left(\prod\limits_{B = 1}^J   d\Pi_{P^B} \right) \rho_{I;J}(R^I|P^J|0) 
\nonumber
\\
&&\,\,\,\,\,\,
\times 
\int_{X^G X^{\prime H}Y^I Y^{\prime J}}
\exp\left\{-\mathrm{i}\big(K^G  X^G
- L^H  X^{\prime H}\big)
+ \mathrm{i}\big(R^IY^I - P^JY^{\prime J}\big)\right\}
\nonumber
\\
&&\,\,\,\,\,\,
\times 
\left(   \prod\limits_{A = 1}^G
\partial_{X^{0,A},E_{K^{A}}} \right)
\left(  \prod\limits_{B = 1}^H
\partial_{X^{0\prime,B},E_{L^B}}^*\right)
\left(\prod\limits_{C = 1}^I \partial_{Y^{0,C},E_{R^C}}^* \right)
\left(\prod\limits_{D = 1}^J 
\partial_{Y^{0\prime,D},E_{P^D}} 
\right)
\nonumber
\\
&&\,\,\,\,\,\,
\times 
\int\mathcal{D}\Phi^{\pm} e^{\mathrm{i}\widehat{S}_{\Phi}[\Phi]}
\Phi^+_{X^G}\Phi^-_{X^{\prime H}}
\exp\left\{\mathrm{i}[\widehat{S}_{\Phi,\text{int}}[\Phi;t] +\widehat{S}_\text{int}[\Phi;t]]\right\}
\Phi^+_{Y^I}\Phi^-_{Y^{\prime J}}
\,\,\,.\,\,\,
\end{eqnarray}
In Eq.\,(\ref{eq:GenDensForm}) we used some short-hand notations from Sec.\,\ref{ssec:Density} and introduced a number new ones. More precisely, we are using
\begin{eqnarray}
\mathrm{i}^G &:=& \mathrm{i}^{g_\alpha + g_\beta + ...}\,\,\,, 
\end{eqnarray}
\begin{eqnarray}
X^{0,G} &:=& x^0_{\alpha,(1)},...,x^0_{\alpha,(g_\alpha)}; x^0_{\beta,(1)},...,x^0_{\beta,(g_\beta)};...\,\,\,,\,\,\,\,\,\,
X^G \,:=\, \mathbf{x}_{\alpha,(1)},...,\mathbf{x}_{\alpha,(g_\alpha)};\mathbf{x}_{\beta,(1)},...,\mathbf{x}_{\beta,(g_\beta)};...\,\,\,, \,\,\,\,\,\,\,\,\,\,\,
\end{eqnarray}
\begin{eqnarray}
K^GX^G &:=& \mathbf{k}_\alpha^{(1)}\mathbf{x}_{\alpha,(1)} + ... + \mathbf{k}_\alpha^{(g_\alpha)}\mathbf{x}_{\alpha,(g_\alpha)} + \mathbf{k}_\beta^{(1)}\mathbf{x}_{\beta,(1)} + ... + \mathbf{k}_\beta^{(g_\beta)}\mathbf{x}_{\beta,(g_\beta)}  +...\,\,\,, 
\end{eqnarray}
\begin{eqnarray}
\prod\limits_{A = 1}^G
\partial_{X^{0,A},E_{K^{A}}} &:=& \partial_{x^{0}_{\alpha,(1)},E^\alpha_{\mathbf{k}_\alpha^{(1)}}} ...\, \partial_{x^{0}_{\alpha,(g_\alpha)},E^\alpha_{\mathbf{k}_\alpha^{(g_\alpha)}}}
\partial_{x^{0}_{\beta,(1)},E^\beta_{\mathbf{k}_\beta^{(1)}}} ...\, \partial_{x^{0}_{\beta,(g_\beta)},E^\beta_{\mathbf{k}_\beta^{(g_\beta)}}}
...\,\,\,, 
\end{eqnarray}
\begin{eqnarray}
\mathcal{D}\Phi^{\pm} &:=& \mathcal{D}\alpha^{\pm}\mathcal{D}\beta^{\pm}...\,\,\,,
\end{eqnarray}
\begin{eqnarray}
\widehat{S}_{\Phi,\text{int}}[\Phi;t] &:=& \widehat{S}_{\alpha,\text{int}}[\alpha;t] + \widehat{S}_{\beta,\text{int}}[\beta;t] + ... \,\,\,,\,\,\,\,\,\,\widehat{S}_\text{int}[\Phi;t] \,=\, \widehat{S}_\text{int}[\alpha;\beta;...;t]+ ...\,\,\,,
\end{eqnarray}
and
\begin{eqnarray}
\Phi^+_{X^G} &:=& \alpha^+_{x_{\alpha,(1)}}...\,\alpha^+_{x_{\alpha,(g_\alpha)}} \beta^+_{x_{\beta,(1)}}...\,\beta^+_{x_{\beta,(g_\beta)}}...
\end{eqnarray}
with
\begin{eqnarray}
\alpha_x &:=& \alpha(x)\,\,\,.
\end{eqnarray}
Taking $G=H=(1,1)$ and assuming all initial density matrix elements except for the $I=J=(1,1)$ one to be vanishing, we recover Eq.\,(\ref{eqn:11DensGen}) from Eq.\,(\ref{eq:GenDensForm}).
\\
It is important to remember that, as was already pointed out in Ref.\,\cite{Kading2022_2}, only the diagonal elements of the $2\times 2$-matrix propagator are permitted when evaluating Eqs.\,(\ref{eqn:11DensGen}) or (\ref{eq:GenDensForm}). This means, when applying Wick's theorem \cite{Wick}, only contractions of fields living on the same branch of the closed time path (cf.\,Fig.\,\ref{fig:CTP}) are possible, such that we only have to work with Feynman and Dyson propagators
\begin{eqnarray}
\bbra{0}  \mathrm{T}[ \hat{\phi}^{+}_x\hat{\phi}^{+}_y ] \kket{0} &=& D^\mathrm{F}_{xy} \,=\, - \mathrm{i}\int_k \frac{e^{\mathrm{i}k\cdot (x-y)}}{k^2+M^2-i\epsilon}\,\,\,,
\\
\bbra{0}  \mathrm{T}[ \hat{\phi}^{-}_x\hat{\phi}^{-}_y ] \kket{0} &=& D^\mathrm{D}_{xy} \,=\, + \mathrm{i}\int_k \frac{e^{\mathrm{i}k\cdot (x-y)}}{k^2+M^2+i\epsilon}\,\,\,,
\\
\bbra{0}  \mathrm{T}[ \hat{\varphi}^{+}_x\hat{\varphi}^{+}_y ] \kket{0} &=& \Delta^\mathrm{F}_{xy} \,=\, - \mathrm{i}\int_k \frac{e^{\mathrm{i}k\cdot (x-y)}}{k^2+m^2-i\epsilon}\,\,\,,
\\
\bbra{0}  \mathrm{T}[ \hat{\varphi}^{-}_x\hat{\varphi}^{-}_y ] \kket{0} &=& \Delta^\mathrm{D}_{xy} \,=\, + \mathrm{i}\int_k \frac{e^{\mathrm{i}k\cdot (x-y)}}{k^2+m^2+i\epsilon}
\,\,\,,
\end{eqnarray}
while the other two-point functions vanish:
\begin{eqnarray}
\bbra{0}\hat{\phi}^{+(-)}_x\hat{\phi}^{-(+)}_y\kket{0}&=& \bbra{0}\hat{\varphi}^{+(-)}_x\hat{\varphi}^{-(+)}_y\kket{0}\,=\, 0\,\,\,.
\end{eqnarray}


\section{Example}
\label{sec:Example}
\begin{figure}[htbp]
\centering
\subfloat[][]{\includegraphics[scale=0.40]{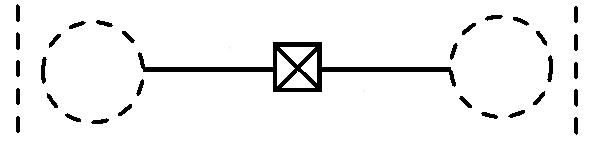}}
\qquad
\subfloat[][]{\includegraphics[scale=0.40]{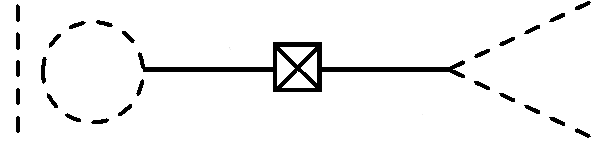}}

\subfloat[][]{\includegraphics[scale=0.40]{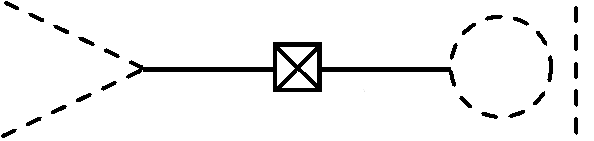}}
\qquad
\subfloat[][]{\includegraphics[scale=0.40]{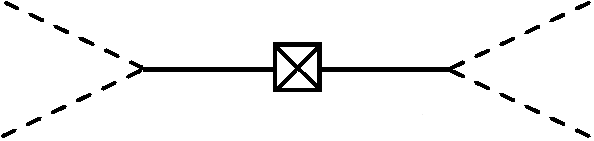}}
\caption{\label{Fig:12} Diagrammatic representations of the terms contributing to $\rho_{0,2;0,2}(t)$, as given in Eq.\,(\ref{eq:12Prop}); the crossed box depicts the insertion of the initial
density matrix element and solid/dashed lines represent $\phi$/$\varphi$-propagators. (a) shows two $\phi$-propagators ending in $\varphi$-tadpoles, while two propagators at equal time but different positions appear. (b) and (c) each depict one $\phi$-
propagator ending in a $\varphi$-tadpole and one propagator at equal time but different positions appearing,
and, in addition, another $\phi$-propagator decaying into two $\varphi$-particles. (d) shows two $\phi$-propagators
each decaying into two $\varphi$-particles.}
\end{figure}

\begin{figure}[htbp]
\centering
\subfloat[][]{\includegraphics[scale=0.40]{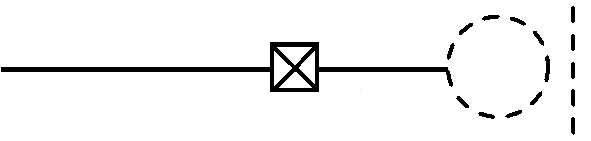}}
\qquad
\subfloat[][]{\includegraphics[scale=0.40]{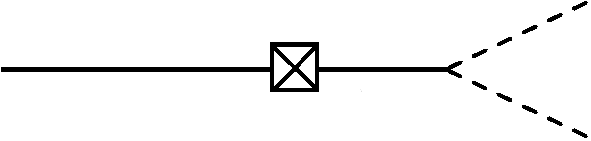}}
\caption{\label{Fig:Partial} Diagrammatic representation of the terms contributing to $\rho_{1,0;0,2}(t)$, as given in Eq.\,(\ref{eq:12Afterpath}); (a) shows one of the $\phi$-propagators ending in a $\varphi$-tadpole, while an equal-time
$\varphi$-propagator is appearing, and another free $\phi$-propagator. (b) contains one $\phi$-propagator decaying
into two $\varphi$-particles and a free $\phi$-propagator. }
\end{figure}

In this section we will apply the formulas in Eqs.\,(\ref{eqn:11DensGen}) and (\ref{eq:GenDensForm}) to a toy model example with two real scalar field species $\phi$ and $\chi$. More precisely, we will consider solely a single particle $\phi$ at the initial time $0$, which is described by the only non-vanishing density matrix $\rho_{1,0;1,0}(0)$.
We will assume that due to an interaction potential $V \sim \varphi^2\phi$ the state of two $\varphi$-particles can become excited within the interval between times $0$ and $t$. This leads to other density matrices potentially being non-vanishing as well at time $t$. Of those we will compute the elements of the matrices $\rho_{0,2;0,2}(t)$, $\rho_{1,0;0,2}(t)$ and $\rho_{1,0;1,0}(t)$. Each of these matrices has its own physical interpretation: the first one corresponds to a $1\!\to 2$-particle decay event, the second describes the correlation between a single $\phi$-particle state and a two $\varphi$-particles state, while the third stands for the single $\phi$-particle state continuing to exist, but with a modified probability. The last of these events, i.e. the change of $\rho_{1,0;1,0}$ over time, can be compared to the dynamics of an open quantum system. 
\\
It should be noted that when computing the density matrices we will sometimes encounter divergences which would usually require renormalization. However, we will only present the unrenormalized expressions here since, as was already mentioned in Ref.\,\cite{Kading2022_2}, the peculiar time-dependent divergences, which we will encounter, have only shortly \cite{Burrage2018} but not yet sufficiently been discussed in the literature. Dealing with such divergences is beyond the scope of the current article and will instead be the subject of a future work \cite{Kading2022}.
\\
We will use the free actions
\begin{eqnarray}
S_\phi[\phi] &=& \int_x \left[ -\frac{1}{2}(\partial\phi)^2 - \frac{1}{2}M^2\phi^2 \right]\,\,\,,
\\
S_\varphi[\varphi] &=& \int_x \left[ -\frac{1}{2}(\partial\varphi)^2 - \frac{1}{2}m^2\varphi^2 \right]
\end{eqnarray}
for the scalar fields $\phi$ and $\varphi$. As interaction action we choose to use
\begin{eqnarray}\label{eq:Interacac}
S_{\text{int}}[\phi;\varphi] &=& \int_{x\in\Omega_t} \left[ -\frac{\alpha}{2}\mathcal{M} \varphi^2\phi \right]\,\,\,,
\end{eqnarray}
where $\alpha \ll 1$ and $\mathcal{M}$ is some sufficiently small mass scale. For simplicity, we do not consider self-interactions. However, including them for more elaborate examples would of course be straightforward, but lead to longer and less readable results.
\\
In every considered case we will have to use the perturbative expansion of the exponentiated interaction action from Eq.\,(\ref{eq:Interacac}). We decide to work up to second order in $\alpha$, such that we find:
\begin{eqnarray}\label{eq:Expon}
\exp\left\{\mathrm{i}\widehat{S}_\text{int}[\phi;\varphi;t]\right\} 
&=& 1 -\mathrm{i}\frac{\alpha}{2}\mathcal{M}\sum\limits_{a=\pm} a \int_z (\varphi_z^a)^2\phi_z^a 
- \frac{\alpha^2}{8}\mathcal{M}^2 \sum\limits_{a,b=\pm} ab \int_{zz'} (\varphi_z^a)^2(\varphi_{z'}^b)^2\phi_z^a\phi_{z'}^b
+ \mathcal{O}(\alpha^3)
\,\,\,,\,\,\,\,\,\,\,\,\,\,\,\,
\end{eqnarray}
where in this and all following equations $z$ and $z'$ are integrated over the domain $\Omega_t$.
We begin our investigations with the decay of a single $\phi$-particle into two $\varphi$-particles, and compute the density matrix elements for this process. Obviously the only non-vanishing density matrix at time $0$ must be $\rho_{1,0;1,0}$, while we want to find the matrix $\rho_{0,2;0,2}$ at time $t$. Following Eq.\,(\ref{eq:GenDensForm}), we must evaluate
\begin{eqnarray}\label{eqn:12Scatt}
\rho_{0,2;0,2}(;\mathbf{p}, \mathbf{k}|;\mathbf{p}', \mathbf{k}'|t) &=&  \lim_{\substack{x^{0(')}_{(1),(2)}\,\to\, t^{+}\\y^{0(')}\,\to\, 0^-}}
\int d\Pi_{\mathbf{q}}d\Pi_{\mathbf{q}'}\rho_{1,0;1,0}(\mathbf{q};|\mathbf{q}';|0)
\nonumber
\\
&&\,\,\,\,\,\,
\times
\int_{\mathbf{x}_{(1)}\mathbf{x}'_{(1)}\mathbf{x}_{(2)}\mathbf{x}'_{(2)}\mathbf{y}\mathbf{y}'} 
e^{-\mathrm{i}(\mathbf{p}\cdot\mathbf{x}_{(1)} + \mathbf{k}\cdot\mathbf{x}_{(2)} - \mathbf{p}'\cdot\mathbf{x}'_{(1)} - \mathbf{k}'\cdot\mathbf{x}'_{(2)})+\mathrm{i}(\mathbf{q}\cdot\mathbf{y}-\mathbf{q}'\cdot\mathbf{y}')}
\nonumber
\\
&&\,\,\,\,\,\,
\times
\partial_{x_{(1)}^0,E^\varphi_{\mathbf{p}}}\partial_{x_{(1)}^{0'},E^\varphi_{\mathbf{p}'}}^*\partial_{x^0_{(2)},E^\varphi_{\mathbf{k}}}\partial_{x^{0'}_{(2)},E^\varphi_{\mathbf{k}'}}^*\partial_{y^0,E^\phi_{\mathbf{q}}}^*
\partial_{y^{0'},E^\phi_{\mathbf{q}'}}
\nonumber
\\
&&\,\,\,\,\,\,
\times
\int\mathcal{D}\phi^{\pm}\mathcal{D}\varphi^{\pm} e^{\mathrm{i}\widehat{S}_{\phi}[\phi]+\mathrm{i}\widehat{S}_{\varphi}[\varphi]}\varphi^+_{x_{(1)}}\varphi^-_{x'_{(1)}}\varphi^+_{x_{(2)}}\varphi^-_{x'_{(2)}}
\exp\left\{\mathrm{i}\widehat{S}_\text{int}[\phi;\varphi;t]\right\}\phi^+_{y}\phi^-_{y'}\,\,\,.
\nonumber
\\
\end{eqnarray}
Substituting Eq.\,(\ref{eq:Expon}), we are left with 
\begin{eqnarray}\label{eq:12Prop}
\rho_{0,2;0,2}(;\mathbf{p}, \mathbf{k}|;\mathbf{p}', \mathbf{k}'|t)
&\approx&
\frac{\alpha^2\mathcal{M}^2}{4}\lim_{\substack{x^{0(')}_{(1),(2)}\,\to\, t^{+}\\y^{0(')}\,\to\, 0^-}}
\int d\Pi_{\mathbf{q}}d\Pi_{\mathbf{q}'}\rho_{1,0;1,0}(\mathbf{q};|\mathbf{q}';|0)
\nonumber
\\
&&\,\,\,\,\,\,
\times
\int_{\mathbf{x}_{(1)}\mathbf{x}'_{(1)}\mathbf{x}_{(2)}\mathbf{x}'_{(2)}\mathbf{y}\mathbf{y}'} 
e^{-\mathrm{i}(\mathbf{p}\cdot\mathbf{x}_{(1)} + \mathbf{k}\cdot\mathbf{x}_{(2)} - \mathbf{p}'\cdot\mathbf{x}'_{(1)} - \mathbf{k}'\cdot\mathbf{x}'_{(2)})+\mathrm{i}(\mathbf{q}\cdot\mathbf{y}-\mathbf{q}'\cdot\mathbf{y}')}
\nonumber
\\
&&\,\,\,\,\,\,
\times
\partial_{x_{(1)}^0,E^\varphi_{\mathbf{p}}}\partial_{x_{(1)}^{0'},E^\varphi_{\mathbf{p}'}}^*\partial_{x^0_{(2)},E^\varphi_{\mathbf{k}}}\partial_{x^{0'}_{(2)},E^\varphi_{\mathbf{k}'}}^*\partial_{y^0,E^\phi_{\mathbf{q}}}^*
\partial_{y^{0'},E^\phi_{\mathbf{q}'}}
\nonumber
\\
&&\,\,\,\,\,\,
\times
\int_{zz'} D_{zy}^\mathrm{F} D_{z'y'}^\mathrm{D}\bigg[   \Delta^{\rm F}_{x_{(1)}x_{(2)}}\Delta^{\rm F}_{zz}  
\Delta^{\rm D}_{x_{(1)}'x_{(2)}'}\Delta^{\rm D}_{z'z'} +2  \Delta^{\rm F}_{x_{(1)}x_{(2)}}\Delta^{\rm F}_{zz}  
\Delta^{\rm D}_{x_{(1)}'z'}\Delta^{\rm D}_{x_{(2)}'z'}
\nonumber
\\
&&\,\,\,\,\,\,\,\,\,\,\,\,\,\,\,\,\,\,
+ 2  \Delta^{\rm F}_{x_{(1)}z}\Delta^{\rm F}_{x_{(2)}z}  
\Delta^{\rm D}_{x_{(1)}'x_{(2)}'}\Delta^{\rm D}_{z'z'}+ 4  \Delta^{\rm F}_{x_{(1)}z}\Delta^{\rm F}_{x_{(2)}z}  
\Delta^{\rm D}_{x_{(1)}'z'}\Delta^{\rm D}_{x_{(2)}'z'}  \bigg]
\,\,\,.
\end{eqnarray}
Using the propagators given in this expression, we can easily depict the contributing physical processes in Figs.\,\ref{Fig:12}(a)-(d). We observe that one (Figs.\,\ref{Fig:12}(b) and (c)) or two (Fig.\,\ref{Fig:12}(a)) of the $\phi$-propagators end in $\varphi$-tadpoles, while, in addition, equally many propagators at equal time but different positions are appearing. Those terms are of course divergent. However, the term diagrammatically represented in Fig.\,\ref{Fig:12}(d) is finite and corresponds to the actual decay of a single $\phi$-particle into two $\varphi$-particles without the appearance of any unphysical terms.
\\
Of course Eq.\,(\ref{eq:12Prop}) can be further evaluated. A brief outline with more details on how to do this can be found in Appendix D of Ref.\,\cite{Kading2019}. Following this procedure, we finally find: 
\begin{eqnarray}\label{eq:r0202result}
\rho_{0,2;0,2}(;\mathbf{p}, \mathbf{k}|;\mathbf{p}', \mathbf{k}'|t)
&\approx&
\frac{\alpha^2 \mathcal{M}^2}{4}
\bigg\{
\frac{\rho_{1,0;1,0}(\mathbf{0};|\mathbf{0};|0)}{M^4}
\delta^2(0)\Delta^{\rm F}_{zz}
\Delta^{\rm D}_{z'z'}
\nonumber
\\
&&\,\,\,\,\,\,
\,\,\,\,\,\,\,\,\,\,\,\,
\times
(2\pi)^6 \delta^{(3)}(\mathbf{p}+\mathbf{k})\delta^{(3)}(\mathbf{p}'+\mathbf{k}')
\sin^2(Mt/2)
\nonumber
\\
&&\,\,\,
\,\,\,\,\,\,\,\,\,\,\,\,
-
\mathrm{i}
\frac{ \rho_{1,0;1,0}(\mathbf{0};|\mathbf{p}'+\mathbf{k}';|0)}{2M^2E^\phi_{\mathbf{p}'+\mathbf{k}'}(E^\phi_{\mathbf{p}'+\mathbf{k}'}-E^\varphi_{\mathbf{p}'} -E^\varphi_{\mathbf{k}'})}
\delta(0)\Delta^{\rm F}_{zz}(2\pi)^3\delta^{(3)}(\mathbf{p}+\mathbf{k})
\nonumber
\\
&&\,\,\,\,\,\,
\,\,\,\,\,\,\,\,\,\,\,\,
\times
\left[ 1 - e^{-\mathrm{i}Mt} \right]
\bigg[ e^{\mathrm{i}(E^\varphi_{\mathbf{p}'} +E^\varphi_{\mathbf{k}'})t}  - e^{\mathrm{i}E^\phi_{\mathbf{p}'+\mathbf{k}'}t} \bigg]
\nonumber
\\
&&\,\,\,
\,\,\,\,\,\,\,\,\,\,\,\,
+
\mathrm{i}
\frac{ \rho_{1,0;1,0}(\mathbf{p}+\mathbf{k};|\mathbf{0};|0)}{2M^2E^\phi_{\mathbf{p}+\mathbf{k}}(E^\phi_{\mathbf{p}+\mathbf{k}}-E^\varphi_{\mathbf{p}} -E^\varphi_{\mathbf{k}})}
\delta(0)\Delta^{\rm D}_{z'z'}(2\pi)^3\delta^{(3)}(\mathbf{p}'+\mathbf{k}')
\nonumber
\\
&&\,\,\,\,\,\,
\,\,\,\,\,\,\,\,\,\,\,\,
\times
\left[ 1 - e^{\mathrm{i}Mt} \right]
\bigg[ e^{-\mathrm{i}(E^\varphi_{\mathbf{p}} +E^\varphi_{\mathbf{k}})t} - e^{-\mathrm{i}E^\phi_{\mathbf{p}+\mathbf{k}}t} \bigg]
\nonumber
\\
&&\,\,\,
\,\,\,\,\,\,\,\,\,\,\,\,
+\frac{\rho_{1,0;1,0}(\mathbf{p}+\mathbf{k};|\mathbf{p}'+\mathbf{k}';|0)}{E^\phi_{\mathbf{p}+\mathbf{k}}E^\phi_{\mathbf{p}'+\mathbf{k}'}(E^\phi_{\mathbf{p}+\mathbf{k}} - E^\varphi_{\mathbf{p}} -E^\varphi_{\mathbf{k}})(E^\phi_{\mathbf{p}'+\mathbf{k}'} - E^\varphi_{\mathbf{p}'} -E^\varphi_{\mathbf{k}'})}
\nonumber
\\
&&\,\,\,\,\,\,
\,\,\,\,\,\,\,\,\,\,\,\,
\times
\bigg[  
e^{-\mathrm{i}(E^\varphi_{\mathbf{p}} + E^\varphi_{\mathbf{k}})t}
-
e^{-\mathrm{i}E^\phi_{\mathbf{p}+\mathbf{k}}t}
\bigg]
\bigg[  
e^{\mathrm{i}(E^\varphi_{\mathbf{p}'} + E^\varphi_{\mathbf{k}'})t}
-
e^{\mathrm{i}E^\phi_{\mathbf{p}'+\mathbf{k}'}t}
\bigg]
\bigg\}
\,\,\,.\,\,\,\,\,\,\,\,\,\,\,\,\,\,\,
\end{eqnarray}
We observe that, as expected, the terms in the fourth, sixth and eighth lines of Eq.\,(\ref{eq:r0202result}) contain differences in the unitary evolutions of a single $\phi$-particle and two $\varphi$-particles. 
\\
Next, we compute one of the two density matrices which describe correlations between a single $\phi$-particle and two $\varphi$-particles. The other one can easily be obtained by making use of the property given in Eq.\,(\ref{eq:DensProperty}). Starting with Eq.\,(\ref{eq:GenDensForm}), we find:
\begin{eqnarray}\label{eqn:partdecay}
\rho_{1,0;0,2}(\mathbf{p}; |;\mathbf{k}',\mathbf{r}'|t) &=& -\mathrm{i} \lim_{\substack{x^{0}_{\phi},x^{0\prime}_{\varphi,(1)},x^{0\prime}_{\varphi,(2)}\,\to\, t^{+}\\y^{0(\prime)}\,\to\, 0^-}}
\int d\Pi_{\mathbf{q}}d\Pi_{\mathbf{q}'}\rho_{1,0;1,0}(\mathbf{q};|\mathbf{q}';|0)
\nonumber
\\
&&\,\,\,\,\,\,
\times
\int_{\mathbf{x}_{\phi}\mathbf{x}'_{\varphi,(1)}\mathbf{x}'_{\varphi,(2)}\mathbf{y}\mathbf{y}'} 
e^{-\mathrm{i}\mathbf{p}\cdot\mathbf{x}_\phi+\mathrm{i}(\mathbf{q}\cdot\mathbf{y}-\mathbf{q}'\cdot\mathbf{y}')}e^{\mathrm{i}(\mathbf{k}'\cdot\mathbf{x}'_{\varphi,(1)}+\mathbf{r}'\cdot\mathbf{x}'_{\varphi,(2)})}
\nonumber
\\
&&\,\,\,\,\,\,
\times
\partial_{x_\phi^0,E^\phi_{\mathbf{p}}}\partial_{y^0,E^\phi_{\mathbf{q}}}^*\partial_{y^{0\prime},E^\phi_{\mathbf{q}'}}
\partial_{x^{0\prime}_{\varphi,(1)},E^\varphi_{\mathbf{k}'}}^*\partial_{x^{0\prime}_{\varphi,(2)},E^\varphi_{\mathbf{r}'}}^*
\nonumber
\\
&&\,\,\,\,\,\,
\times
\int\mathcal{D}\phi^{\pm}\mathcal{D}\varphi^{\pm} e^{\mathrm{i}\widehat{S}_{\phi}[\phi]+\mathrm{i}\widehat{S}_{\varphi}[\varphi]}\phi^+_{x_\phi}\varphi^-_{x'_{\varphi,(1)}}\varphi^-_{x'_{\varphi,(2)}}
\exp\left\{\mathrm{i}\widehat{S}_\text{int}[\phi;\varphi;t]\right\}\phi^+_{y}\phi^-_{y'}
\,\,\,.\,\,\,\,\,\,\,\,\,\,\,\,
\end{eqnarray}
Evaluating the path integrals leads us to
\begin{eqnarray}\label{eq:12Afterpath}
\rho_{1,0;0,2}(\mathbf{p}; |;\mathbf{k}',\mathbf{r}'|t)
&\approx&
\frac{\alpha\mathcal{M}}{2}
\lim_{\substack{x^{0}_{\phi},x^{0\prime}_{\varphi,(1)},x^{0\prime}_{\varphi,(2)}\,\to\, t^{+}\\y^{0(\prime)}\,\to\, 0^-}}
\int d\Pi_{\mathbf{q}}d\Pi_{\mathbf{q}'}\rho_{1,0;1,0}(\mathbf{q};|\mathbf{q}';|0)
\nonumber
\\
&&\,\,\,\,\,\,
\times
\int_{\mathbf{x}_{\phi}\mathbf{x}'_{\varphi,(1)}\mathbf{x}'_{\varphi,(2)}\mathbf{y}\mathbf{y}'} 
e^{-\mathrm{i}\mathbf{p}\cdot\mathbf{x}_\phi+\mathrm{i}(\mathbf{q}\cdot\mathbf{y}-\mathbf{q}'\cdot\mathbf{y}')}e^{\mathrm{i}(\mathbf{k}'\cdot\mathbf{x}'_{\varphi,(1)}+\mathbf{r}'\cdot\mathbf{x}'_{\varphi,(2)})}
\nonumber
\\
&&\,\,\,\,\,\,
\times
\partial_{x_\phi^0,E^\phi_{\mathbf{p}}}\partial_{y^0,E^\phi_{\mathbf{q}}}^*\partial_{y^{0\prime},E^\phi_{\mathbf{q}'}}
\partial_{x^{0\prime}_{\varphi,(1)},E^\varphi_{\mathbf{k}'}}^*\partial_{x^{0\prime}_{\varphi,(2)},E^\varphi_{\mathbf{r}'}}^*
\nonumber
\\
&&\,\,\,\,\,\,
\times
\int_z
D^{\rm F}_{x_\phi y} D^{\rm D}_{z y'}
\bigg[ \Delta^{\rm D}_{x'_{\varphi,(1)} x'_{\varphi,(2)}} \Delta^{\rm D}_{zz} + 2  \Delta^{\rm D}_{x'_{\varphi,(1)} z} \Delta^{\rm D}_{x'_{\varphi,(2)} z}\bigg]
\,\,\,,\,\,\,\,\,\,\,\,\,\,\,\,
\end{eqnarray}
which can be used to depict the processes as in Figs.\,\ref{Fig:Partial}(a) and (b). As in the previous case, we observe one of the $\phi$-propagators ending in a $\varphi$-tadpole, while an equal-time $\varphi$-propagator is appearing, in Fig.\,\ref{Fig:Partial}(a). In contrast, Fig.\,\ref{Fig:Partial}(b) contains only physical terms without any divergences.
\\
Continuing the computation, we find:
\begin{eqnarray}
\rho_{1,0;0,2}(\mathbf{p}; |;\mathbf{k}',\mathbf{r}'|t)
&\approx&
\frac{\alpha\mathcal{M}e^{-\mathrm{i}E^\phi_\mathbf{p}t}}{2}
\bigg\{
-\mathrm{i}\delta(0) \Delta^{\rm D}_{zz}(2\pi)^3\delta^{(3)}(\mathbf{k}'+\mathbf{r}')\frac{\rho_{1,0;1,0}(\mathbf{p};|\mathbf{0};|0)}{2M^2} \left[ 1 - e^{\mathrm{i}Mt} \right]
\nonumber
\\
&&
\,\,\,\,\,\,\,\,\,\,\,\,\,\,\,\,\,\,\,\,\,\,\,\,\,\,\,\,\,\,
+
\frac{\rho_{1,0;1,0}(\mathbf{p};|\mathbf{k}'+\mathbf{r}';|0)}{E^\phi_{\mathbf{k}'+\mathbf{r}'}(E^\phi_{\mathbf{k}'+\mathbf{r}'} - E^\varphi_{\mathbf{k}'} -E^\varphi_{\mathbf{r}'})}
\left[  
e^{\mathrm{i}E^\phi_{\mathbf{k}'+\mathbf{r}'}t}
-
e^{\mathrm{i}(E^\varphi_{\mathbf{k}'} + E^\varphi_{\mathbf{r}'})t}
\right]
\bigg\}
\,\,\,,\,\,\,\,\,\,\,\,\,\,\,\,\,\,\,
\end{eqnarray}
where, in the last line, we again find the difference in the unitary evolutions of a single $\phi$-particle and two $\varphi$-particles.
\\
Finally, we study the change of $\rho_{1,0;1,0}$ over time. We use Eq.\,(\ref{eq:GenDensForm}) to find:

\begin{eqnarray}\label{eqn:deco}
\rho_{1,0;1,0}(\mathbf{p};|\mathbf{p}'; |t) &=&  \lim_{\substack{x^{0(\prime)}\,\to\, t^{+}\\y^{0(\prime)}\,\to\, 0^-}}
\int d\Pi_{\mathbf{q}}d\Pi_{\mathbf{q}'}\rho_{1,0;1,0}(\mathbf{q};|\mathbf{q}';|0)
\int_{\mathbf{x}\mathbf{x}'\mathbf{y}\mathbf{y}'} 
e^{-\mathrm{i}(\mathbf{p}\cdot\mathbf{x} - \mathbf{p}'\cdot\mathbf{x}') + \mathrm{i}(\mathbf{q}\cdot\mathbf{y} - \mathbf{q}'\cdot\mathbf{y}') }
\nonumber
\\
&&\,\,\,\,\,\,
\times
\partial_{x^0,E^\phi_{\mathbf{p}}}\partial_{x^{0'},E^\phi_{\mathbf{p}'}}^*
\partial_{y^{0},E^\phi_{\mathbf{q}}}^*\partial_{y^{0\prime},E^\phi_{\mathbf{q}'}}
\int\mathcal{D}\phi^{\pm}\mathcal{D}\varphi^{\pm} e^{\mathrm{i}\widehat{S}_{\phi}[\phi]+\mathrm{i}\widehat{S}_{\varphi}[\varphi]}\phi^+_{x}\phi^-_{x'}
\nonumber
\\
&&\,\,\,\,\,\,
\times
\exp\left\{\mathrm{i}\widehat{S}_\text{int}[\phi;\varphi;t]\right\}\phi^+_{y}\phi^-_{y'}
\,\,\,,
\end{eqnarray}
which becomes
\begin{eqnarray}\label{eq:1010}
\rho_{1,0;1,0}(\mathbf{p};|\mathbf{p}'; |t)
&\approx&  \lim_{\substack{x^{0(\prime)}\,\to\, t^{+}\\y^{0(\prime)}\,\to\, 0^-}}
\int d\Pi_{\mathbf{q}}d\Pi_{\mathbf{q}'}\rho_{1,0;1,0}(\mathbf{q};|\mathbf{q}';|0)
\int_{\mathbf{x}\mathbf{x}'\mathbf{y}\mathbf{y}'} 
e^{-\mathrm{i}(\mathbf{p}\cdot\mathbf{x} - \mathbf{p}'\cdot\mathbf{x}') + \mathrm{i}(\mathbf{q}\cdot\mathbf{y} - \mathbf{q}'\cdot\mathbf{y}') }
\nonumber
\\
&&\,\,\,\,\,\,
\times
\partial_{x^0,E^\phi_{\mathbf{p}}}\partial_{x^{0'},E^\phi_{\mathbf{p}'}}^*
\partial_{y^{0},E^\phi_{\mathbf{q}}}^*\partial_{y^{0\prime},E^\phi_{\mathbf{q}'}}
\bigg\{ [1+\aleph(t)] D^{\rm F}_{xy}D^{\rm D}_{x'y'} 
\nonumber
\\
&&\,\,\,\,\,\,
-\frac{\alpha^2}{4}\mathcal{M}^2 \int_{zz'} \bigg[
\Big( \Delta^{\rm F}_{zz}\Delta^{\rm F}_{z'z'} + 2 (\Delta^{\rm D}_{zz'})^2 \Big) D^{\rm F}_{xy} D^{\rm D}_{x'z} D^{\rm D}_{z'y'}
\nonumber
\\
&&\,\,\,\,\,\,\,\,\,\,\,\,\,\,\,\,\,\,\,\,\,\,\,\,\,\,\,\,\,\,\,\,\,\,\,\,\,\,\,\,
+  (x,y\longleftrightarrow x',y')^\ast
\bigg]
\bigg\}
\end{eqnarray}
after evaluating the path integrals. Here we introduced
\begin{eqnarray}\label{eq:Aleph}
\aleph(t)
&:=&
- \frac{\alpha^2}{8}\mathcal{M}^2 \sum\limits_{a=\text{F},\text{D}}  \int_{zz'} \left[ \Delta^{a}_{zz}\Delta^{a}_{z'z'} + 2 (\Delta^{a}_{zz'})^2 \right]D^{a}_{zz'}
\nonumber
\\
&\approx& 
-\frac{\alpha^2\mathcal{M}^2}{2}
 (2\pi)^3\delta^{(3)}(\mathbf{0})\left\{
 \frac{\sin^2(Mt/2)}{M^3} \Delta^{\rm F}_{zz}\Delta^{\rm F}_{z'z'}
+\frac{1}{2}\int_{\mathbf{q}\mathbf{l}}  \frac{\sin^2[(E^\phi_\mathbf{q}+E^\varphi_\mathbf{l}+E^\varphi_{\mathbf{q}+\mathbf{l}})t/2]}{E^\phi_\mathbf{q}E^\varphi_\mathbf{l}E^\varphi_{\mathbf{q}+\mathbf{l}}(E^\phi_\mathbf{q}+E^\varphi_\mathbf{l}+E^\varphi_{\mathbf{q}+\mathbf{l}})^2}
\right\}
\,\,\,,\,\,\,\,\,\,\,\,\,\,\,\,\,
\end{eqnarray}
which only contains disconnected loop diagrams. Interestingly, at this stage, Eq.\,(\ref{eq:1010}) is exactly the same as the equation obtained for the time evolution of the single-particle reduced density matrix elements in Ref.\,\cite{Kading2022_2}. Therefore, we point to the results and figures presented in this reference. The reasons for the formalism presented here giving the same results as the one in Ref.\,\cite{Kading2022_2} lie in us only working up to second order in $\alpha$ and in our choice of the interaction and self-interaction potentials. Even though the formalism in Ref.\,\cite{Kading2022_2} generally permits contractions for some $+$-fields with $-$-fields (the ones representing the traced-out environmental degrees of freedom), the usage of the interaction potential $V \sim \varphi^2\phi$ (Ref.\,\cite{Kading2022_2} names the $\varphi$-field $\chi$ instead) does not allow for them to appear in this particular computation to order $\alpha^2$. It is important to note that considering higher orders and/or including self-interactions would lead, in addition to Feynman and Dyson propagators, to the appearance of Wightman propagators for the $\chi$-fields in the computation in Ref.\,\cite{Kading2022_2}. A corresponding computation in this article, on the other hand, would lead to the appearance of Feynman and Dyson propagators only and consequently to different final results.


\section{Conclusions and outlook}
\label{sec:Conclusion}

Density matrices are powerful tools in non-relativistic quantum mechanics, but also in quantum field theory. However, analytically solving the quantum master equations that describe the time evolution of density matrices can be extremely complicated or even hopeless.
\\
In this article we used the ideas originally introduced in Refs.\,\cite{Burrage2018} and \cite{Kading2022_2} for developing a practicable and first principle-based density matrix formalism for interacting quantum fields. It has the advantage of circumventing quantum master equations by providing a relatively simple formula for directly computing density matrices in a momentum basis for any number of field species and occupations in Fock space.
\\
As a working example, we chose the setup of two different real scalar fields interacting with each other, but also provided an equation that allows for the inclusion of arbitrarily more field species. Extending the formalism to other types of fields, e.g.\,\,complex scalar fields, vector fields etc., is rather straightforward. Before deriving the relevant formulas, we discussed density matrices in Fock space, see Sec.\,\ref{ssec:Density}, the Schwinger-Keldysh formalism, Sec.\,\ref{ssec:FVIF}, and thermo field dynamics, Sec.\,\ref{ssec:TFD}. Those were the ingredients required for the actual derivation, which we concluded in Sec.\,\ref{ssec:EQN}. Finally, in Sec.\,\ref{sec:Example}, we applied the developed formalism to a selected example.
\\
While the formalism in Ref.\,\cite{Kading2022_2} is specifically designed for open quantum systems and therefore includes the tracing-out of environmental degrees of freedom, the formalism presented here captures all degrees of freedom equally without loss of any information. This is very useful, for example, for studies of creation and annihilation processes at finite times, including particle decays. Of course the formalism presented in this article also naturally leads to similar time-dependent divergences as in Ref.\,\cite{Kading2022_2}, which will be discussed in more detail in a future work \cite{Kading2022}. However, while in Ref.\,\cite{Kading2022_2} every computation led to the emergence of such divergences due to the process of tracing out the environmental degrees of freedom, in this article  we encountered only a few of those at the considered order in perturbation theory since we allowed for external legs of all involved scalar fields to appear. 
\\
All in all, the present work is an important first step towards a description of physically relevant interacting quantum field theory systems at finite times. In particular, we see imminent applications of the presented formalism in a variety of areas at the high-precision frontier, i.e. in the area of low-energy but high-precision phenomenology as, e.g. in quantum optics experiments.
 

\begin{acknowledgments}

We thank T.~Colas and P.~Millington for useful comments and helpful discussions.
This work was supported by the Austrian Science Fund (FWF): P 34240-N.
\end{acknowledgments}


\bibliography{DMsfFV}
\bibliographystyle{JHEP}

\end{document}